\documentclass[aps,prl,twocolumn,amsmath,amssymb,showpacs,superscriptaddress,notitlepage,longbibliography,10pt]{revtex4-2}

\usepackage[colorlinks=true,linkcolor=blue,anchorcolor=red,citecolor=blue, urlcolor=blue]{hyperref}
\usepackage{bm}
\usepackage{graphicx}
\usepackage{physics}
\usepackage{color}
\usepackage{multirow}
\usepackage{extarrows}
\usepackage{array}
\usepackage{appendix}

\usepackage{cases}
\usepackage[normalem]{ulem}
\def\Fig#1{\text{Fig.}~\ref{#1}}
\def\Eq#1{Eq.~(\ref{#1})}

\begin{document}
\title{3D Quantum Hall Effect with Two Distinct Plateaus}

\author{Jun-Hong Li$^\dag$}
\affiliation{State Key Laboratory of Quantum Functional Materials, Department of Physics, and Guangdong Basic Research Center of Excellence for Quantum Science, Southern University of Science and Technology (SUSTech), Shenzhen 518055, China}

\author{Yi-Yuan Chen$^\dag$}
\affiliation{Quantum Science Center of Guangdong-Hong Kong-Macao Greater Bay Area (Guangdong), Shenzhen 518045, China}

\author{Peng-Lu Zhao}
\email{Corresponding author: zhaopl@gmail.com}
\affiliation{Quantum Science Center of Guangdong-Hong Kong-Macao Greater Bay Area (Guangdong), Shenzhen 518045, China}

\author{Hai-Zhou Lu}
\email{Corresponding author: luhz@sustech.edu.cn}
\affiliation{State Key Laboratory of Quantum Functional Materials, Department of Physics, and Guangdong Basic Research Center of Excellence for Quantum Science, Southern University of Science and Technology (SUSTech), Shenzhen 518055, China}
\affiliation{Quantum Science Center of Guangdong-Hong Kong-Macao Greater Bay Area (Guangdong), Shenzhen 518045, China}

\author{X. C. Xie}
\affiliation{International Center for Quantum Materials, School of Physics, Peking University, Beijing 100871, China}
\affiliation{Interdisciplinary Center for Theoretical Physics and Information
Sciences (ICTPIS), Fudan University, Shanghai 200433, China}
\affiliation{Hefei National Laboratory, Hefei 230088, China}

\begin{abstract}
The recent discovery of the 3D quantum Hall effect in $\mathrm{HfTe_5}$ has also revealed puzzling signatures of possible 3D fractionalization. Beyond the first plateau associated with the lowest Landau band, Hall conductivity exhibits a second plateau with a value of about $3/5$ of the first, accompanied by a suppressed longitudinal resistivity. Here, we attribute this second plateau to an insulating ground state arising from spin-density-wave order. We show that a magnetic-field-driven Lifshitz transition causes the spin-down holelike zeroth Landau band to cross the Fermi energy and that the resulting nesting between the lowest spin-up and spin-down Landau bands induces a spin-density wave. We calculate the Hall and longitudinal resistivity and reproduce the experimental behaviors. Our renormalization-group analysis further supports this insulating ground state. Our work reveals that the tunability of Landau bands along the magnetic-field direction endows the 3D quantum Hall effect with a broader phenomenology than its 2D counterpart and merits further exploration.
 
\end{abstract}
\maketitle

\begin{figure}[t] 
    \centering    
    \includegraphics[width=0.48\textwidth]{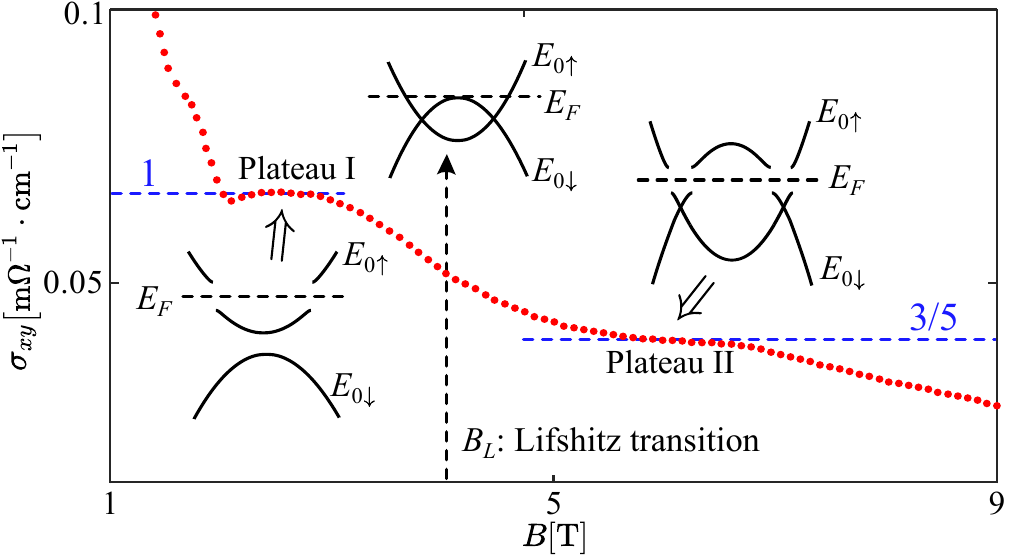} 
    \caption{Summary of two distinct plateaus in the 3D quantum Hall effect of HfTe$_5$. The red dots are experimental Hall-conductivity data, adapted from Ref.~\cite{galeski2020unconventional}. The blue dashed lines indicate the first plateau and the second one, whose value is $3/5$ of the first. The insets show the evolution of the lowest Landau bands with increasing $B$: the first plateau is associated with a charge-density-wave gap in the $E_{0\uparrow}$ band before the Lifshitz transition, whereas beyond $B_L$ the interband nesting between $E_{0\uparrow}$ and $E_{0\downarrow}$ favors a spin-density wave and produces the second plateau.}\label{FSum} 
\end{figure}

\textcolor{blue}{\emph{Introduction}}--The quantum Hall effect is a cornerstone of condensed-matter physics \cite{klitzing1980new,tsui1982two,laughlin1983anomalous,thouless1982quantized}. 
Conventionally, it arises from Landau quantization in a 2D electron gas: when the Fermi energy lies between adjacent Landau levels, the bulk becomes insulating while edge states support dissipationless and quantized transport. This mechanism makes the quantum Hall effect inherently a 2D phenomenon \cite{zhang2005experimental}. By contrast, in 3D systems, electrons remain dispersive along the magnetic-field direction, making it difficult to form the insulating bulk required for quantized Hall transport. As a result, realizing a quantum Hall effect in 3D has remained a longstanding challenge \cite{lu20193d,halperin1987possible,montambaux1990quantized,kohmoto1992diophantine,koshino2001hofstadter,bernevig2007theory,stormer1986quantization,cooper1989quantized,hannahs1989quantum,hill1998bulk,cao2012quantized,masuda2016quantum,liu2016zeeman,wang20173d,zhang2017room,uchida2017quantum,schumann2018observation,zhang2019quantum,liu2021spin,li20203d,cheng2020quantum}. Recently, quantized Hall resistivity and conductivity were observed in 3D devices of $\mathrm{ZrTe_5}$ \cite{tang2019three,Galeski2021} and $\mathrm{HfTe_5}$ \cite{galeski2020unconventional,PhysRevB.101.161201}, respectively, providing compelling evidence for a 3D quantum Hall effect. In these systems, a density wave formed from Landau bands has been proposed to gap the bulk and enable quantized Hall transport \cite{halperin1987possible,qin2020theory}, as illustrated in \Fig{FSum}. More intriguingly, $\mathrm{HfTe_5}$ exhibits an additional Hall plateau beyond the first one, with a Hall conductivity approximately $3/5$ of the first plateau (see \Fig{FSum}) and accompanied by a reduced longitudinal resistivity. Whether this additional plateau reflects a 3D fractional quantum Hall effect, and more generally the microscopic origin of the two distinct plateaus in the 3D quantum Hall effect, remains unclear.

In this Letter, we show that the second plateau in the 3D quantum Hall effect originates from a spin-density wave that opens a global gap in the ultraquantum limit, as summarized in \Fig{FSum}. We further demonstrate that this spin-density wave becomes possible after a Lifshitz transition, where the spin-down zeroth Landau band crosses the Fermi energy and generates four Fermi points, enabling an interband nesting instability. Using band parameters appropriate for HfTe$_5$, we find that Hall conductivity and longitudinal resistivity calculated with a spin-density-wave gap are in good agreement with experiment \cite{galeski2020unconventional}. Our renormalization-group analysis shows that this instability is driven by electron-phonon coupling through a Peierls mechanism. These results uncover a new mechanism for the 3D quantum Hall effect and reveal how Fermi-surface reconstruction and interaction-driven gap opening cooperate to produce emergent quantum Hall phenomena beyond the 2D paradigm.
\begin{figure}[t] 
    \centering
    \includegraphics[width=0.48\textwidth]{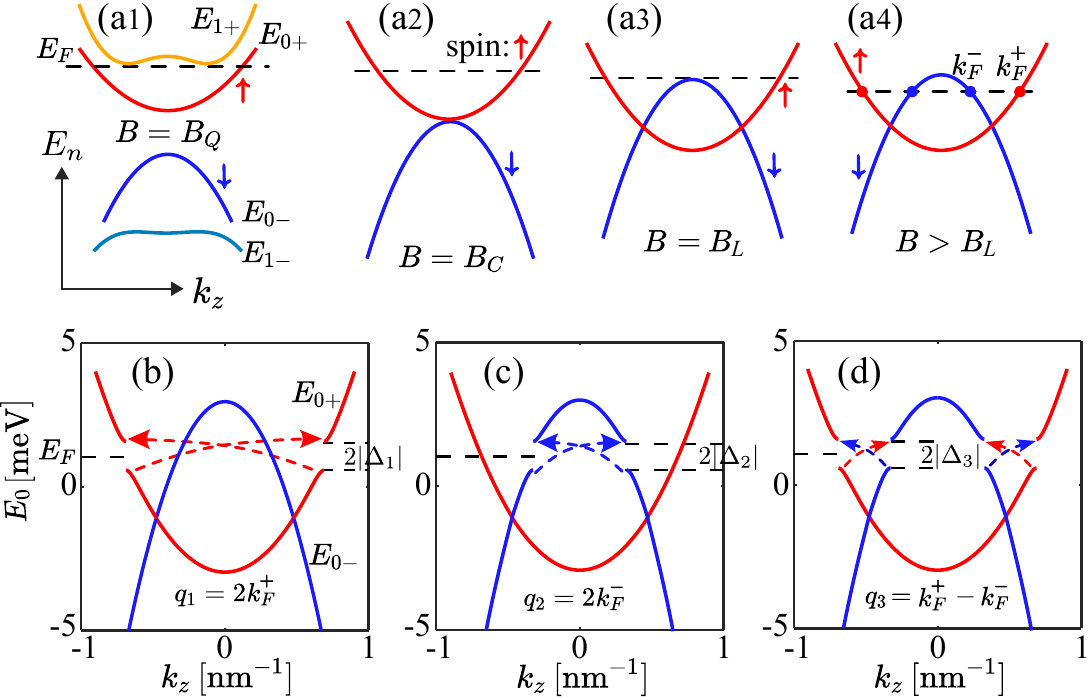} 
    \caption{(a1)–(a4) Evolution of the 0th Landau bands with increasing $B$. Here, $+$ and $-$ label spin up (red $\uparrow$) and spin down (blue $\downarrow$). (a1) The quantum limit is reached at $B=B_Q$. (a2) The 0th Landau bands cross at $B=B_C$. (a3) The Lifshitz transition occurs at $B=B_L$. (a4) For $B>B_L$, $E_{0\pm}$ cross $E_F$ and form four Fermi points, $\pm k_F^{\pm}$. (b)–(d) Three distinct gap openings, $2|\Delta_{1,2,3}|$, due to three Fermi-surface nesting channels: (b),(c) intraband nesting within $E_{0+}$ and $E_{0-}$, giving charge-density waves with $q_1=2k_F^{+}$ and $q_2=2k_F^{-}$, respectively; (d) interband nesting between $E_{0+}$ and $E_{0-}$, giving a spin-density wave with $q_3=k_F^{+}-k_F^{-}$.}
\label{FBandEvo}
\end{figure}

\textcolor{blue}{\emph{Lifshitz transition}}--We show that the spin-down zeroth Landau band in a weak topological insulator moves upward with increasing magnetic field and crosses the Fermi energy, driving a Lifshitz transition (see \Fig{FBandEvo}(a1)–(a4)). We start with a generic Dirac model $\left(\hslash=c=1\right)$\cite{shen17b,chen2015magnetoinfrared}
\begin{align} 
    \mathcal{H}_0(\mathbf{k})= &D_zk_z^2+ v_{x} k_{x} \tau_{x}  \sigma_{z}+v_{y} k_{y} \tau_{y}  \sigma_{0}+v_{z} k_{z} \tau_{x}  \sigma_{x}  \nonumber \\
    & +\left[M_{0}+M_{1}\left(v_{x}^{2} k_{x}^{2}+v_{y}^{2} k_{y}^{2}\right)+M_{z} k_{z}^{2}\right] \tau_{z} \sigma_{0}, \label{H_e}
\end{align}
where $v_{x,y,z}$ denote the Fermi velocities, $\tau_{x,y,z,0}$ and $\sigma_{x,y,z,0}$ are the Pauli and identity matrices acting on the orbital and spin degrees of freedom, respectively, $2|M_{0}|$ sets the bulk gap size, and $M_{1,z}$ are parameters characterizing the band inversion. This model describes semimetals, strong and weak topological insulators, and normal insulators, depending on the values of $M_{0}$, $M_{1}$, and $M_{z}$  \cite{Wenghm14PRX,fan2017transition,manzoni2016evidence,chen2015optical,nair2018thermodynamic,li2016chiral,li2016experimental,chen2017spectroscopic,jiang2017landau,xiong2017three,xu2018temperature}. Here we focus on the weak topological insulator phase with $M_0M_1<0$ and $M_0M_z>0$ \cite{wu2023topological,ZPL25PRM,ZJL25PRL,Jauregui25PRL}.

In a uniform magnetic field $\mathbf{B} = B\mathbf{e}_z$, the 3D band dispersion of this model reduces to a series of 1D Landau bands, whose explicit forms can be found in Refs.~\cite{wu2023topological,ZPL25PRM}. We will focus on the quantum limit, in which the Fermi energy $E_F$ crosses only the zeroth Landau bands (see \Fig{FBandEvo}(a1)) with dispersions $E_{0\pm}=D_zk_z^2\pm(M_0+M_1v_xv_y eB+M_zk_z^2)$. Without loss of generality, we take $M_0>0, M_1<0$, and $M_z>0$. As the magnetic field increases, $E_{0+}$ moves downward while $E_{0-}$ moves upward, and at a critical field $B_L$ the $E_{0-}$ band crosses the Fermi energy, producing a Lifshitz transition (\Fig{FBandEvo}(a3)). Strong experimental evidence for this Lifshitz transition has been reported in HfTe$_5$~\cite{wu2023topological,Galeski22NC,ZJL25PRL}. We thus adopt the band parameters of this material as $M_0=9.02\ \mathrm{meV}$, $M_1v_xv_y=-1.58\ \mathrm{eV}\cdot\mathrm{nm^2}$, $M_z=13.1\ \mathrm{meV}\cdot\mathrm{nm^2}$, $D_z=-4.58\ \mathrm{meV}\cdot\mathrm{nm^2}$, and use a carrier density $n_0=1.7\times 10^{17}\ \mathrm{cm}^{-3}$. Consequently, the system reaches the quantum limit at the magnetic field $B_Q=1.8\ \mathrm{T}$~\cite{galeski2020unconventional,qin2020theory}, and the Lifshitz transition occurs at magnetic field $B_L=4.2\ \mathrm{T}$ \cite{wu2023topological,ZJL25PRL,ZPL25PRM}.

\textcolor{blue}{\emph{Spin density wave}}--In the following, we analyze Fermi-surface instabilities within mean field \cite{gruner2018density,giamarchi2003quantum}. We show that electron–phonon–induced particle–hole pairing between the two zeroth Landau bands 
generates the spin-density wave in \Fig{FBandEvo}(d) and opens a global gap.

As shown in \Fig{FSum}(a4), the Fermi surface consists of four Fermi points at $rk_{F}^{\mu}$, where $\mu=\pm$ labels the Landau bands $E_{0\mu}$ crossed by the Fermi energy and $r=\mp$ denotes the two crossings. Due to Fermi-surface nesting \cite{gruner2018density,giamarchi2003quantum}, the Lindhard response function $\chi_{\mu\nu}(q)=(2\pi)^{-1}\int\mathrm{d}k_z (f_{k_z}-f_{k_z+q})/(E_{0\mu}(k_z)-E_{0\nu}(k_z+q))$ diverges at the three wave vectors $q_{1,2,3}=2k_{F}^{+},2k_{F}^{-}$, and $k_{F}^{+}-k_{F}^{-}$ \cite{mahan2013many}, where $f$ is the Fermi-Dirac distribution. These nesting wave vectors allow electrons and holes near the Fermi points to pair and develop three types of density-wave order. The corresponding order parameters, $\Delta_{q}^{\mu\nu} \propto \sum_{\delta k_z}\left\langle\hat{c}_{\delta k_z+r k_F^{\mu}+q }^{\dagger} \hat{c}_{r' k_F^{\nu}+\delta k_z}\right\rangle$ with $q=|r k_F^{\mu}-r'k_F^{\nu}|=q_{1,2,3}$, involve $\hat{c}_{\mathbf{k}}^\dagger$ and $\hat{c}_{\mathbf{k}}$, the creation and annihilation operators for electrons with wave vector $\mathbf{k}$. Here, the particle–hole pair is located near the Fermi points $r k_F^{\mu}$ and $r' k_F^{\nu}$, $\delta k_z=k_z-r k_F^{\mu}$ describes the deviation from the Fermi points. As shown in \Fig{FBandEvo}(b) and (c), the cases $q=q_1$ with $\mu=\nu=+$ and $q=q_2$ with $\mu=\nu=-$, correspond to charge-density waves of spin-up and spin-down electrons, respectively. Neither of these two cases opens a global gap, and the system remains metallic. \Fig{FBandEvo}(d) shows the case $q=q_3$ with $\mu\nu=-$, yielding a spin-density wave and opening a global gap. The mean-field analysis presented below identifies the spin-density wave in \Fig{FBandEvo}(d) as the leading instability in strong magnetic fields, as it alone remains appreciable at high fields.

In the presence of electron–phonon interactions, Fermi-surface nesting translates into a lattice distortion. The order parameters correspond to phonon modes with finite expectation values at the nesting wave vectors, and the mean-field Hamiltonian is given by \cite{gruner1988dynamics,bruus2004many,qin2020theory}
\begin{align}
  \mathcal{H}_{m} = &\sum_{\delta k_z}\sum_{\mu,r=\pm} \hslash v_{F}^{\mu r} \delta k_z \hat{c}_{\delta k_z +rk_{F}^{\mu}}^{\dagger} \hat{c}_{\delta k_z+rk_{F}^{\mu}} \nonumber
    \\ & +
    \sum_{\delta k_z}|\Delta_{q}^{\mu\nu}|\left(\mathrm{e}^{i \phi} \hat{c}_{\delta k_z +rk_{F}^{\mu}}^{\dagger} \hat{c}_{\delta k_z+r'k_{F}^{\nu}}+\text{H.c.}\right)
    \nonumber \\ & +\frac{|\Delta_{q}^{\mu\nu}|^2}{g_{q}},  \label{EqH_t}
    \end{align}
where $\hslash v_{F}^{\mu r}=2(D_z+\mu M_z) r k_{F}^{\mu}$, $g_q = g_0/(q^2+\kappa^2)^2$ is the screened Yukawa potential mediated by ions, with $g_0$ an electron-phonon coupling constant and $1/\kappa$ the screening length obtained following \cite{qin2020theory}. 
The first line in \Eq{EqH_t} is the Hamiltonian of free-electrons obtained by linearizing $E_{0\mu}$ at Fermi points. The second and third lines describe the density wave order in the mean-field induced by electron–phonon coupling; the absence of a sum over $\mu$ or $r$ indicates that only a density wave $q=q_1$, $q_2$ or $q_3$ is retained, without coexistence of multiple orders. For each $q$ (and the corresponding indices $\mu,\nu$), we diagonalize \Eq{EqH_t} separately to obtain the quasiparticle spectrum of density-waves $E_q\left(k_z \right)$. The order parameter $\Delta_{q}^{\mu\nu}$ is determined by minimizing the total energy $E_g$ of electrons and phonons, i.e., $\partial E_g/\partial|\Delta_{q}^{\mu\nu}|=0$, with
\begin{equation}
    E_g=\sum_\mathbf{k} [E_q\left(k_z \right)-E_F]f(E_q\left(k_z \right))+\frac{|\Delta_{q}^{\mu\nu}|^2V}{g_q},
\end{equation}
where $V$ is the volume.

\begin{figure}[t] 
    \centering
    \includegraphics[width=0.47\textwidth]{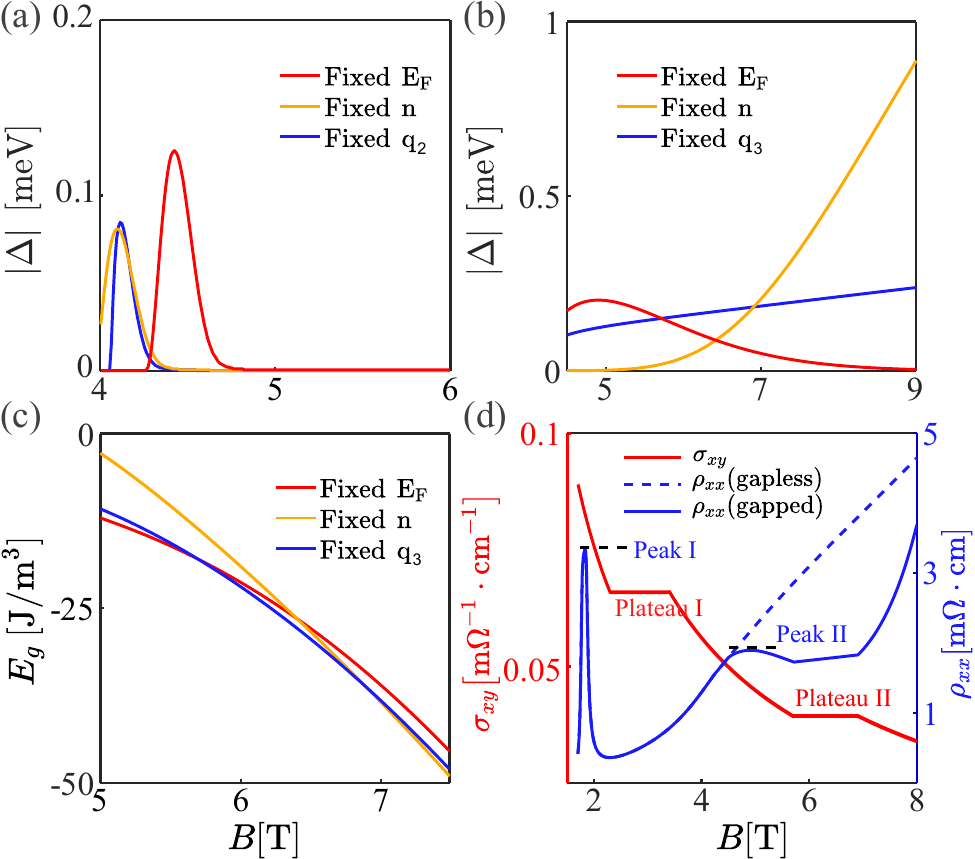} 
    \caption{(a),(b) Calculated $|\Delta(B)|$ for density waves with wave vectors $q_2$ and $q_3$, respectively. In both cases, the calculations are performed for three cases: fixed Fermi energy $E_F=0.7~\mathrm{meV}$, fixed carrier density $n=1.7\times10^{17}~\mathrm{cm}^{-3}$, and fixed nesting wave vector $q=0.32~\mathrm{n m}^{-1}$ [i.e., fixed $q_2$ in (a) and fixed $q_3$ in (b)]. (c) Ground-state energy corresponding to the gap shown in (b), evaluated for the same three cases. (d) Calculated $\sigma_{xy}$ and $\rho_{xx}$ over $1.7\text{--}8~\mathrm{T}$. The blue solid curve shows $\rho_{xx}$ calculated with gap opening above $4.5~\mathrm{T}$, while the blue dashed curve denotes $\rho_{xx}$ calculated without gap opening over the entire field range; below $4.5~\mathrm{T}$, the dashed curve is obscured by the solid one. The band parameters used here are given above in the discussion of the Lifshitz transition; in addition, $v_x=0.45~\mathrm{eV\cdot nm}$, $v_y=0.2~\mathrm{eV\cdot nm}$~\cite{tang2019three,wu2023topological}, $g_0=268~\mathrm{eV\cdot nm^{-1}}$, and $\epsilon=8.5$~\cite{Mahin2026}. For the dashed $\rho_{xx}(B)$ curve in (d), calculated following Ref.~\cite{ZPL25PRM}, we take $\Gamma_1=0.5~\mathrm{meV}$ and $\Gamma_0=5~\mathrm{meV}$. 
     }
\label{fig:conductivity}
\end{figure}

\Fig{fig:conductivity}(a) and \Fig{fig:conductivity}(b) show the calculated $|\Delta_q^{\mu\nu}|$ for a charge-density wave with wave vector $q_2$ and a spin-density wave with wave vector $q_3$, respectively. For each $q$, we evaluate $|\Delta|$ under three assumptions: (i) a fixed, field-independent $E_F$, (ii) a fixed carrier density $n$, and (iii) a fixed $q$. The validity of these assumptions has been discussed in Refs.~\cite{qin2020theory,ZJL25PRL,ZPL25PRM}. A nonzero $|\Delta|$ signals the onset of these orders. The results for $q_1$ are omitted, since $|\Delta|$ remains zero throughout this range of $B$, indicating that no order with wave vector $q_1$ develops. \Fig{fig:conductivity}(a) shows that the charge-density-wave gap ($q=q_2$) becomes nearly negligible for $B>5~\mathrm{T}$. By contrast, the spin-density-wave gap at $q=q_3$ remains appreciable in \Fig{fig:conductivity}(b), indicating that the spin-density wave is the leading Fermi-surface instability in strong magnetic fields. Moreover, the charge-density wave gaps out only the $E_{0+}$ band, while the $E_{0-}$ band remains gapless [\Fig{FBandEvo}(c)]. The bulk state therefore remains metallic. In comparison, the spin-density wave opens a global gap [\Fig{FBandEvo}(d)] and drives a metal-insulator transition, which may account for the metal-insulator transition observed in HfTe$_5$ \cite{galeski2020unconventional,PhysRevB.101.161201,Jauregui25PRL} over a similar range of $B$.

\textcolor{blue}{\emph{Transport Signatures}}--We now show that the opening of the spin-density-wave gap gives rise to a field-independent plateau in the Hall conductivity over $5.7\text{--}6.9~\mathrm{T}$ and a peak in the longitudinal resistivity at $4.9~\mathrm{T}$, as indicated by the red and blue curves in \Fig{fig:conductivity}(d), respectively. The coexistence of these two anomalous transport features is highly unusual and constitutes a clear hallmark of gap opening.

\Fig{fig:conductivity}(c) presents the corresponding ground-state energy $E_g$. Notably, the solid blue curve shows that $E_g$ is minimized over a broad field range ($5.7\text{--}6.9~\mathrm{T}$) when $q_3$ is held constant. Consequently, the 3D Hall conductivity becomes field independent, since
\begin{equation}\label{sigmaxy}
    \sigma_{xy}^{3D}=\frac{e^2}{h}\frac{k_F^{+}-k_F^{-}}{\pi}.
\end{equation}
The above expression follows because each fixed-$k_z$ Landau band contributes $e^2/h$ in the quantum limit, and summing over occupied $k_z$ states yields $(k_F^{+}-k_F^{-})/\pi$. As shown by the red line in \Fig{fig:conductivity}(d), the Hall conductivity is calculated for $B \in (1.7, 8.0)\,\mathrm{T}$. A first Hall plateau appears over $2.3\text{--}3.4~\mathrm{T}$. This plateau has been observed in both ZrTe$_5$ and HfTe$_5$ \cite{tang2019three,galeski2020unconventional,PhysRevB.101.161201,Galeski2021} and is commonly interpreted as a typical manifestation of the 3D quantum Hall effect. Its origin has been attributed to a charge-density wave formed in the $E_{0+}$ band near the Fermi surface before the Lifshitz transition \cite{qin2020theory}. For $2.8<B<4.5~\mathrm{T}$, the charge-density wave vanishes, while the spin-density wave has not yet formed. The system therefore remains gapless, and $\sigma_{xy}=en_0/B\propto B^{-1}$, where the carrier density is given by $n_0=eB(k_F^{+}-k_F^{-})/(\pi h)$ \cite{ZPL25PRM}. The spin-density-wave gap opens above $4.5~\mathrm{T}$, with $|\Delta(B)|$ shown in \Fig{fig:conductivity}(b). Among the three cases in \Fig{fig:conductivity}(b), the largest gap yields the lowest energy and thus determines the true gap. For $4.5<B<5.7~\mathrm{T}$, the gap develops with pinned Fermi energy, so the Hall conductivity decreases with increasing $B$. For $5.7<B<6.9~\mathrm{T}$, $k_F^{+}-k_F^{-}$ is fixed, giving rise to a Hall plateau. We take $q_3=k_F^{+}-k_F^{-}=0.32~\mathrm{nm}^{-1}$, which yields a plateau height comparable to the measured value in \cite{galeski2020unconventional}, $\sigma_{xy}=3e^2k_{Fz}/(5\pi h)$ with $k_{Fz}=0.58~\mathrm{nm}^{-1}$ extracted from quantum oscillations. The plateau range also depends sensitively on $q_3$, which is also close to the experimental value reported in \cite{galeski2020unconventional}. The plateau disappears at $B=6.9~\mathrm{T}$, where the gap obtained at fixed density becomes the largest. Beyond this field, $\sigma_{xy}$ again follows $e\tilde{n}/B\propto B^{-1}$, and no further plateau appears.

The calculated longitudinal resistivity $\rho_{xx}$ also reflects gap opening. As shown by the blue curve in \Fig{fig:conductivity}(c), $\rho_{xx}$ exhibits a peak at the quantum limit, $B_{\mathrm{Q}}=1.8~\mathrm{T}$, and then decreases rapidly. This behavior follows from the Kubo-formula calculation \cite{ZPL25PRM,ZJL25PRL}, which applies to $B<4.5~\mathrm{T}$, i.e., before the gap opens. After gap opening, $\rho_{xx}$ follows the Arrhenius form, $\rho_{xx}=\rho_0 \exp(|\Delta|/k_B T)$ \cite{Jauregui25PRL}, producing a dome-like feature over $4.5\text{--}5.7~\mathrm{T}$ that mainly tracks the dome-like field dependence of $|\Delta|$ in \Fig{fig:conductivity}(b) for fixed $E_F$. By contrast, the gaps at fixed $q_3$ and fixed $n$ both increase with $B$. As a result, $\rho_{xx}$ develops an additional peak around $4.9~\mathrm{T}$ in addition to the one at $B_{\mathrm{Q}}$; this peak would be absent without gap opening. The dashed blue curve in \Fig{fig:conductivity}(c) shows $\rho_{xx}$ calculated without a gap \cite{ZPL25PRM,ZJL25PRL}, which increases monotonically with $B$ in the deep quantum limit. This additional peak provides another transport signature of the spin-density wave. Such a peak was also observed in \cite{galeski2020unconventional}; moreover, the calculated blue curve in \Fig{fig:conductivity}(c) remains in qualitative agreement with the experimental data over the entire field range.

\textcolor{blue}{\emph{Renormalization-group analysis}}--To confirm that the global gap is induced by electron--phonon, rather than electron--electron, interactions, we perform a Wilsonian momentum-shell renormalization-group analysis \cite{fisher1974renormalization,wilson1983renormalization,mackinnon1994critical,stanley1999scaling,wilson1974renormalization,abrikosov2012methods,peskin2018introduction,shankar1994renormalization}. The momentum shell is defined as $\Lambda e^{-\ell}<|k_z|<\Lambda$, where $\ell$ is the running scale parameter. The free action follows from the effective Hamiltonian obtained by linearizing the dispersion near $k_{F}^{+}$ and $k_{F}^{-}$,
\begin{equation}\label{S0}
S_0=\int\mathrm{d}^3x\,\mathrm{d}\tau\,\psi^\dagger\left[\partial_{\tau}-\mathrm{i}\partial_{z}\left(v_{-}\sigma_0+v_{+}\sigma_z\right)\right]\psi,
\end{equation}
where $\psi=(\psi_-,\psi_+)^{\mathrm{T}}$ with $\psi_\pm$ annihilating (and $\psi_\pm^\dagger$ creating) fermions near $k_F^\pm$, and $v_\pm=(M_z +D_z)k_F^{+}\pm (M_z -D_z)k_F^{-}$.

We consider a system with electron–phonon, electron–electron interactions and back-scattering disorder:
\begin{align}
S_{\mathrm{int} }=&\int\mathrm{d}^3x\mathrm{d}\tau\left(\mathcal{L}_{ep}+\mathcal{L}_{b}+\mathcal{L}_{e}\right) +\int\mathrm{d}^3x\mathrm{d}\tau\mathrm{d}\tau'\mathcal{L}_d, \nonumber \\
    \mathcal{L}_{ep}=&g\left(\phi\psi_+^\dagger\psi_-+\phi^*\psi_-^\dagger\psi_+\right), \nonumber \\
     \mathcal{L}_{p}=&\left|\partial_\tau\phi\right|^2+v_p^2\left|\partial_k\phi\right|^2,\quad
    \mathcal{L}_{ee}=u\psi_+^\dagger\psi_+\psi_-^\dagger\psi_+, \nonumber \\
      \mathcal{L}_d=& -\frac{\Delta_b}{2}\left(\psi^\dagger\Gamma_b\psi\right)_\tau\left(\psi^\dagger\Gamma_b\psi\right)_{\tau'}. \label{S}
\end{align}
Here $\mathcal{L}_{ep}$ describes electron--phonon interactions at the mean-field level and $\mathcal{L}_{ee}$ encodes electron--electron interactions, both leading to the $q_3$ instability ($q_3= k_F^{+}-k_F^{-}$). $\mathcal{L}_{p}$ describes the order-parameter dynamics \cite{Sachdev_2011}, with $r=0$ at the quantum critical point. $\mathcal{L}_d$ is the disorder Lagrangian after disorder averaging using the replica method. For backscattering disorder, $\Gamma_b=\sigma_x+\sigma_y$ scatters electrons between $k_F^{+}$ and $k_F^{-}$ \cite{giamarchi2003quantum,giamarchi1988anderson}.
\begin{figure} [t]
    \centering
    \includegraphics[width=0.46\textwidth]{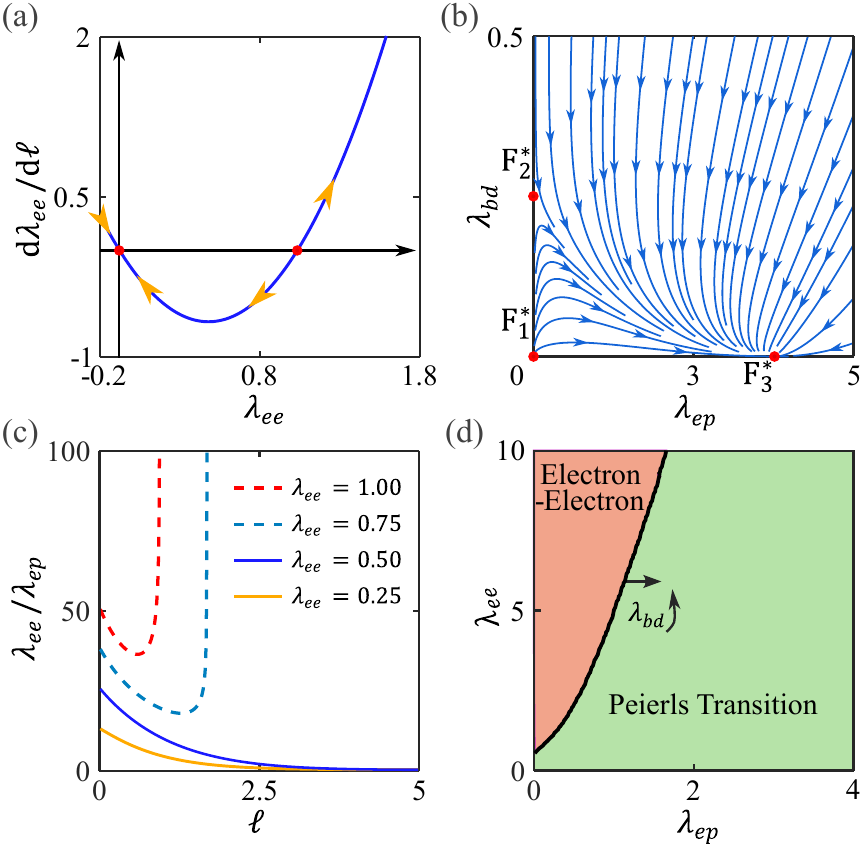}
    \caption{(a) Schematic plot of $\mathrm{d}\lambda_{ee}/\mathrm{d}\ell$ in \Eq{RG equations} versus $\lambda_{ee}$. The arrows indicate the renormalization-group flow near the fixed points. (b) Renormalization-group flow in the $\lambda_{ep}$-$\lambda_{bd}$ plane at $\lambda_{ee}^{\ast}=0$, calculated from \Eq{RG equations} with $C_1=0.43$, $C_2=0.89$, and $C_3=2.11$. (c) Evolution of $\lambda_{ee}/\lambda_{ep}$ with $\ell$, obtained for $\lambda_{ep}(\ell=0)=0.02$, $\lambda_{bd}(\ell=0)=0.5$, with four initial values of $\lambda_{ee}$ indicated in the legend. (d) Phase diagram in the $\lambda_{ep}$-$\lambda_{ee}$ plane, obtained from \Eq{RG equations} with $\lambda_{bd}(\ell=0)=1$. The black curve denotes the phase boundary between the electron-electron-driven and Peierls-transition-driven phases. The arrow labeled $\lambda_{bd}$ indicates that, as $\lambda_{bd}(\ell=0)$ increases, the phase boundary shifts to the right.}
\label{fig:RG} 
\end{figure}

Based on the renormalization-group calculations, we introduce three effective coupling strengths: $\lambda_{ep}$ for electron--phonon interactions, $\lambda_{ee}$ for electron--electron interactions, and $\lambda_{bd}$ for backscattering disorder, defined as $\lambda_{ep}\equiv g^2/(4\pi^2\ell_B^2 v_p^3\Lambda^2)$, $\lambda_{ee}\equiv u/(4\pi^2 v_{+}\ell_B^2)$, and $\lambda_{bd}\equiv \Delta_b/[2\pi^2 (v_{+}^2-v_{-}^2)\ell_B^2\Lambda]$, whose renormalization-group equations are given by
\begin{align}
\frac{\mathrm{d}v_{\pm}}{\mathrm{d}\ell}&=(z_f-1+C_1\lambda_{ep}-C_3\lambda_{bd})v_{\pm},\label{RG equations} \\
\frac{\mathrm{d}\lambda_{ee}}{\mathrm{d}\ell}&=2\lambda_{ee}^2-(C_2\lambda_{ep}+C_3\lambda_{bd})\lambda_{ee}, \nonumber \\
\frac{\mathrm{d}\lambda_{ep}}{\mathrm{d}\ell}&=(C_1-C_2)\lambda_{ep}^2+2(1+\lambda_{ee}-C_3\lambda_{bd})\lambda_{ep},\nonumber \\
\frac{\mathrm{d}\lambda_{bd}}{\mathrm{d}\ell}&=-4\lambda_{bd}^2+[1+2\lambda_{ee}-(C_1+C_2)\lambda_{ep}] \lambda_{bd}, \nonumber
\end{align}
where the fermionic dynamical exponent $z_f(\ell)=1-C_1\lambda_{ep}(\ell)+C_3\lambda_{bd}(\ell)$ is obtained by requiring $v_{\pm}$ to be $\ell$-independent, and $C_i$ ($i=1,2,3$) are constants that depend on $\eta_f\equiv v_{-}/v_{+}$ and $\eta_b\equiv v_{p}/v_{+}$:
$C_1=2\eta_b^3\left[(\eta_b+1)^2+\eta_f^2\right]/\left[(\eta_b+1)^2-\eta_f^2\right]^2$, $C_2=2\eta_b^2(\eta_b+1)/[(\eta_b+1)^2-\eta_f^2]$,
and $C_3=2(\eta_f^2+1)/(1-\eta_f^2)$.

By setting the flows in \Eq{RG equations} to zero, we obtain five nonnegative fixed points, $F^*_{1-5}$, located at $(\lambda_{ee},\lambda_{ep},\lambda_{bd})=(0,0,0)$, $(0,2/(C_2-C_1),0)$, $(0,0,1/4)$, $(C_3/(8-2C_3),0,1/(4-C_3))$, and $\big([3C_3(C_1+C_2)-8C_2]/8C_1,(3C_3-8)/4C_1,3/4\big)$, respectively. The value of $C_3$ is crucial for identifying the nonnegative fixed points of \Eq{RG equations}. For the band parameters used in \Fig{fig:conductivity}, $C_3$ lies in the range $2\text{--}2.24$ for $4.5\mathrm{T}<B<9\mathrm{T}$. Therefore, among the seven fixed points, only the five listed above remain nonnegative.

As illustrated in \Fig{fig:RG}(a), the fixed points with $\lambda_{ee}>0$ are unstable; namely, $\mathrm{F^*_{4}}$ and $\mathrm{F^*_{5}}$ are both unstable. These two fixed points satisfy $\lambda_{ee}=(C_2\lambda_{ep}+C_3\lambda_{bd})/2$ and thus lie at the critical point for $\lambda_{ee}$. For $\lambda_{ee}<(C_2\lambda_{ep}+C_3\lambda_{bd})/2$, $\lambda_{ee}$ flows to zero and the corresponding fixed points are $\mathrm{F^*_{1-3}}$. The flow diagram among $\mathrm{F^*_{1}}$, $\mathrm{F^*_{2}}$, and $\mathrm{F^*_{3}}$ is presented in \Fig{fig:RG}(b), which shows that $\mathrm{F^*_{2}}$ is the stable fixed point. In this regime, the renormalization-group flows approach $\lambda_{ee}/\lambda_{ep}\to 0$ (see solid lines in \Fig{fig:RG}(c)), indicating that the spin-density wave is formed via a Peierls transition. By contrast, when $\lambda_{ee}>(C_2\lambda_{ep}+C_3\lambda_{bd})/2$, $\lambda_{ee}$, $\lambda_{ep}$, and $\lambda_{bd}$ flow to $\infty$. At the same time, $\lambda_{ee}/\lambda_{ep}\to \infty$, as depicted by the dashed lines in \Fig{fig:RG}(c). This divergence implies that the spin-density wave is formed by the electron-electron interactions. \Fig{fig:RG}(d) presents the boundary between these two microscopic origins in the $(\lambda_{ep},\lambda_{ee})$ plane at fixed $\lambda_{bd}$. The arrow denotes a shift of the boundary to the right as $\lambda_{bd}$ increases, suggesting that backscattering disorder suppresses the Peierls transition and favors the electron-electron-induced transition.

\textcolor{blue}{\emph{Discussion}}--Using the band parameters of HfTe$_5$, the calculated Hall conductivity and longitudinal resistivity in \Fig{fig:conductivity}(d) agree well with the measurements in \cite{galeski2020unconventional}, providing strong support for our theory. In principle, our theory applies to all weak topological insulators. The practical issue is whether the magnetic field required to observe the Lifshitz transition lies within the experimentally accessible limit. Previous results show that the critical field for the Lifshitz transition $B_{L}\propto n_0^{3/2}$\cite{wu2023topological,ZJL25PRL}. Therefore, weak topological insulators with low carrier density are ideal materials for observing this new quantum Hall effects. However, the second plateau is generally difficult to realize in strong topological insulators, where the conditions for a Lifshitz transition are very restrictive. This may explain why only the first Hall plateau has been observed in ZrTe$_5$ \cite{tang2019three,galeski2020unconventional}, but not the second. Earlier experiments also demonstrated that strain can drive ZrTe$_5$ from a strong topological insulator to a weak one \cite{fan2017transition,comphy22,scirep17}. It would be very interesting to see whether the second Hall plateau also emerges in ZrTe$_5$ under such tuning.

\textcolor{blue}{\emph{Acknowledgments}}--We thank helpful discussions with Xiangang Wan and Xiaoqun Wang. This work was supported by the National Key R$\&$D Program of China (2022YFA1403700), Innovation Program for Quantum Science and Technology (2021ZD0302400), the National Natural Science Foundation of China (Grants No. 12304074, 12525401, and 12350402), Guangdong Basic and Applied Basic Research Foundation (2023B0303000011), Guangdong Provincial Quantum Science Strategic Initiative (GDZX2201001 and GDZX2401001), Guangdong province (2020KCXTD001), the Science, Technology and Innovation Commission of Shenzhen Municipality (ZDSYS20190902092905285), and Center for Computational Science and Engineering of SUSTech. 

$^\dag$J.-H.L. and Y.-Y.C. contributed equally.

\bibliography{ref}

\end{document}